\documentclass[preprint]{aastex}
\usepackage{emulateapj5}
\usepackage{onecolfloat5}
\usepackage{apjfonts}
\usepackage{epsfig}

\def\plotone#1{\centering \leavevmode
\epsfxsize= 1.0\columnwidth \epsfbox{#1}}

\def\gsim{\;\rlap{\lower 2.5pt
 \hbox{$\sim$}}\raise 1.5pt\hbox{$>$}\;}
\def\lsim{\;\rlap{\lower 2.5pt
   \hbox{$\sim$}}\raise 1.5pt\hbox{$<$}\;}
\newcommand{\lya}{Ly$\alpha$ }
\newcommand{\be}{\begin{equation}}
\newcommand{\beq}{\begin{equation}}
\newcommand{\ba}{\begin{eqnarray}}
\newcommand{\ee}{\end{equation}}
\newcommand{\eeq}{\end{equation}}
\newcommand{\ea}{\end{eqnarray}}
\newcommand{\etal}{{et al.}}
\newcommand{\msun}{$M_{\odot}$}

\begin{document}
\twocolumn[
\submitted{Submitted to ApJ}
\title{Implications of the Ly$\alpha$ Emission Line from a Candidate z=10 Galaxy}

\author{Renyue Cen}
\affil{Department of Astrophysical Sciences, Princeton University, Peyton Hall, Ivy Lane, Princeton, NJ 08544}

\vspace{-0.5\baselineskip} 
\author{Zolt\'{a}n Haiman, and Andrei Mesinger}
\affil{Department of Astronomy, Columbia University, 550 West 120th Street, New York, NY 10027}
              
\begin{abstract}
The recently discovered $z=10$ galaxy (Pello \etal\ 2004) has a strong
Ly$\alpha$ emission line that is consistent with being surprisingly symmetric, even
given the relatively poor quality of its spectrum.  The blue wing of a
Ly$\alpha$ line originating at high redshift should be strongly
suppressed by resonant hydrogen absorption along the line of sight, an
expectation borne out by the observed asymmetric shapes of the
existing sample of \lya\ emitting sources at lower redshifts ($3< z <
6.7$).  Absorption on the blue side of the line of the Pello \etal\
source could be reduced if the intergalactic medium (IGM) in the
vicinity of the galaxy is highly ionized, but we show that this
requires an unrealistically high ionizing emissivity.  We suggest
instead that the Ly$\alpha$ emitting gas be receding relative to the
surrounding gas with a velocity of $\gsim 35$ km/s, a large 
velocity that is plausible only if the galaxy is part of a larger
system (group of galaxies) with a velocity dispersion $\gsim 35$ km/s.
We find that with this velocity shift, the observed strength and shape
of the line is still consistent with the galaxy being surrounded
by its own Str\"omgren sphere embedded in a
fully neutral IGM.  More generally, we predict that at any given
redshift, the bright Ly$\alpha$ emitters with broader lines would
exhibit stronger asymmetry than fainter ones.  Bright galaxies with
symmetric Ly$\alpha$ lines may be signposts for groups and clusters of
galaxies, within which they can acquire random velocities comparable
to or larger than their linewidths.
\end{abstract}]

\section{Introduction}
\label{sec:introduction}

Three independent observations combine to paint a complex picture of
the cosmological reionization process.  First, the recent quasar
absorption spectrum observations by the Sloan Digital Sky Survey show
strong evidence that the reionization process completes at $z\sim 6$
(Becker \etal\ 2001; Fan \etal\ 2002; Cen \& McDonald 2002).
Second, the latest {\it Wilkinson Microwave Anisotropy Probe (WMAP)}
observations detect a high Thomson scattering optical depth,
suggesting that the intergalactic medium (IGM) experienced a
significantly ionized state at high redshift somewhere between
$z=15-25$ (Kogut \etal\ 2003) for (at least) a significant redshift
interval.  This is somewhat contradicted by the third observational
line of evidence of the intergalactic medium having a relatively high
temperature at $z\sim 3-4$, which requires a reionization epoch no
earlier than redshift $z=9-10$ (Hui \& Haiman 2003; Theuns \etal\
2002).  While the overall picture is consistent with a pre-WMAP,
physically motivated double reionization model (Cen 2003; Wyithe \&
Loeb 2003), a detailed probe of the ionization state of the IGM at
high redshift is sorely wanted.

Ly$\alpha$ emission lines from high--redshift sources can serve as
probes of the ionization state of the IGM. The damping wing of the
Gunn--Peterson (GP) absorption from the IGM can cause a characteristic
absorption feature (Miralda-Escud\'e 1998).  For a Ly$\alpha$ emitting
galaxy embedded in a partly neutral IGM, the absorption produces
conspicuous effects, i.e. attenuating the emission line, making it
asymmetric, and shifting its apparent peak to longer wavelengths
(Haiman 2002; Santos 2003).  In practice, the expectation is that
strong conclusions cannot be drawn from a single galaxy.  For example,
the relatively strong Ly$\alpha$ line of the $z=6.6$ galaxy discovered
by Hu \etal (2002) is still consistent with being embedded in a
neutral IGM, but surrounded by its own ionized Str\"omgren sphere
(Haiman 2002; Santos 2003).

The recent claim of the detection of a Ly$\alpha$ emitting galaxy at
$z=10$ (Pello \etal\ 2004; hereafter P04) provides a new opportunity
to study the IGM at $z=10$.  This source is especially interesting,
since at its high inferred redshift, absorption by the IGM should
increase significantly.  In this {\it Letter}, we examine both the
observed shape and overall attenuation of the detected Ly$\alpha$
line, in models where the line is processed through the IGM. We find
that in order to achieve the observed symmetry of the Ly$\alpha$ line
profile, the emitting gas in the galaxy must be receding faster than
the surrounding gas by at least $35$km/s.  Given this required
recessional velocity, we find that the P04 source is marginally
consistent with being embedded in a fully neutral IGM at $z=10$, with
the line suffering an attenuation by a factor of 30.
 
Throughout this paper, we assume the background cosmology to be flat
$\Lambda$CDM with $(\Omega_\Lambda,\Omega_{\rm m},\Omega_{\rm b},
h)=(0.7,0.3,0.04,0.7)$.

\section{The Observed Ly$\alpha$ Emission Line}

We first consider a characterization of the symmetry of the observed
Ly$\alpha$ emission line.  The general expectation, based on simple
models of absorption, is that the blue side of the line should be
strongly suppressed relative to the red side (Haiman 2002; Santos
2003); this expectation is borne out by \lya\ emitting sources
detected at lower redshifts, nearly all of which show such asymmetry
when at high enough spectral S/N (Rhoads et al. 2003; Shapley et
al. 2003; Trager et al. 1997).  In contrast, Figure 5a of P04 reveals
that the line is nearly symmetric, with perhaps more flux on its {\it
blue} side.  Our goal here is to quantify the significance of the
apparent lack of the expected asymmetry.  We define a flux ratio,
$R\equiv L_b/L_r$, as a measure of the line symmetry, where $L_b$ and
$L_r$ are the total flux in the Ly$\alpha$ line blueward and redward
of its {\it apparent} peak wavelength, respectively.

In general, consider the spectrum of an emission line with significant
flux detected in $N=N_b+N_r$ pixels ($N_b$ and $N_r$ being the number
of pixels on the blue/red side of the line, respectively). The signal
(flux) in each pixel is given by $b_1, b_2, ... b_{N_b}$ and $r_1,
r_2, ... r_{N_r}$, with associated noise $\sigma(b_1)$,
$\sigma(b_2)...$, and $\sigma(r_1)$, $\sigma(r_2)...$.  In this case,
we can obtain the mean value and the uncertainty of the line asymmetry
parameter $R$ by usual error propagation as follows:

\beq
\langle R \rangle = \frac{b_1 + b_2 + ... b_{N_b}}{r_1 + r_2 + ... r_{N_r}}
\eeq

\noindent and

\begin{eqnarray}
\sigma(R)^2 &=&
 \frac{\sigma(b_1)^2 + \sigma(b_2)^2 + ... \sigma(b_{N_b})^2}{(r_1 + r_2 + ... r_{N_r})^2} +
\\
&&  [\sigma(r_1)^2 + \sigma(r_2)^2 + ... \sigma(r_{N_r})]^2
 \frac{(b_1 + b_2 + ... b_{N_b})^2}{(r_1 + r_2 + ... r_{N_r})^4}.
\end{eqnarray}

In the emission line spectrum displayed in Figure 5a of P04, there are
six pixels with $S/N>1$, with the apparent line center defined to lie
at the center of the central pixel.  We therefore break the central
pixel into half, and consider $b_1=0.35, b_2=0.75, b_3=1.25$, and
$b_4=0.8$, and $r_1=0.8, r_2=1.4$, and $r_3=0.6$, where the fluxes are
quoted for pixels in order of increasing wavelength, and in units of
$10^{-18}~{\rm erg~s^{-1}~cm^{-2}}$~\AA$^{-1}$.  We take the noise to
be a constant in each full pixel
$\sigma(b_1)=\sigma(b_2)=\sigma(b_3)=\sigma(r_2)=\sigma(r_3)=0.2$, and
we lower its value by a factor of $\sqrt{2}$ for the central
half-pixels, $\sigma(b_4)=\sigma(r_1)=0.14$.  We find

\beq
R=1.12\pm 0.05.
\eeq

\noindent 
Assuming the flux in each pixel is a normally distributed random
variable, we find that the line significantly close to being
symmetric, and, in fact, appears to have an asymmetry with more flux
on the {\it blue} side of the line, which is the opposite of the
expectations.  For the sake of concreteness, in the rest of this
paper, we will require $0.9<R<1.3$. Taking the data at face value,
this corresponds to a $\sim 4\sigma$ statistical confidence limit on
this ratio.  However, we note that the observed line is only
marginally resolved, with its width only approximately $40\%$ larger
than the spectral resolution (P04). We briefly examine the extent to
which convolving the emission line with the instrumental response
(approximately described by a 60km/s Gaussian) reduces the asymmetry.
As an example, we adopt a Ly$\alpha$ line, assumed to be described by
a Gaussian with a width of $50$km/s, but with the blue flux suppressed
by a factor of $\exp(-a\Delta\lambda^2)$. Here $\Delta\lambda$ is the
offset from the Lyman $\alpha$ wavelength, and $a$ is a constant
chosen such that the line has an asymmetry of $R=0.90$. We find that
when this line is convolved with the PSF, the asymmetry is degraded
and somewhat symmetrized to $R=0.92$.  However, we also find that the
effect of the convolution depends strongly on the otherwise poorly
known intrinsic line width and shape. Thus, in order to decide
whether the emission line of the P04 galaxy candidate is indeed as
symmetric as it appears at the current, low spectral S/N and
resolution, it would be highly desirable to repeat this type of
analysis with a higher quality spectrum, when available.
 
\section{Transmission of the \lya Emission Line}
\label{sec:transmission}

An ionizing source embedded in the high--redshift IGM will maintain an
ionized region around it (Str\"omgren sphere).  We solve the equation
of motion for the ionization front exactly (Shapiro \& Giroux 1987;
Cen \& Haiman 2000; Haiman 2002) in an evolving density field, taking
into account recombinations.  The optical depth between the source and
the observer at $z=0$, at the observed wavelength $\lambda_{\rm
obs}=\lambda_s(1+z_s)$, is given by $\tau(\lambda_{\rm
obs},z_s)=\int_{z_r}^{z_s} dz c \frac{dt}{dz} n_H(z)
\sigma_\alpha[\lambda_{\rm obs}/(1+z)]$, where $cdt/dz$ is the line
element in the assumed $\Lambda$CDM cosmology, $n_H$ is the neutral
hydrogen density, and $\sigma_\alpha$ is the Ly$\alpha$ absorption
cross--section at $\lambda_{\rm obs}/(1+z)$.  There are two separate
contributions to the optical depth from within ($z_i < z < z_s$) and
outside ($z_r < z < z_i$) the Str\"omgren sphere, where $z_i$ is the
redshift somewhat below $z_s$, corresponding to the boundary of the
Str\"omgren sphere ($z_i\approx z_s-R_s/R_H[z_s]$, where $R_H[z_s]$ is
the size of the cosmological horizon at $z_s$).  A numerically
computed Voigt profile (see eq.~6 of Press \& Rybicki 1993) is used
for $\sigma_\alpha$.  Outside the Str\"omgren sphere the IGM is
assumed to have a neutral fraction $x_{\rm HI}$ with mean density.

Inside the HII region, the neutral hydrogen density is calculated
assuming photoionization equilibrium, with a gas temperature of
$T=10^4$K.  We assume a log-normal distribution for the gas density and
compute the effective opacity $\tau_{\rm eff}=-\ln \langle
\exp(-\tau_{\rm HI})\rangle$, where the brackets denote averaging over
the probability distribution of $\tau_{\rm HI}$.  We also adopt an
ionizing emissivity of the source as described in the next section. We
find that our results are insensitive to the choice of clumping
(defined as $C\equiv\langle n_H^2\rangle/\langle n_H\rangle^2$) in the
range $1<C<1000$ due to two competing factors. On one hand, a larger
$C$ gives a larger fraction of low optical depth regions for the
assumed density distribution. On the other hand, a larger $C$ results
in a larger overall neutral fraction in ionization equilibrium.
Therefore, the adoption of the log-normal form for the density
distribution is not critical for our results.  Nevertheless, to
explicitly check the validity of the assumption of a log--normal
density PDF, we derive the PDF directly from a hydrodynamic simulation
(see Cen \etal\ 2004 for details).  We find that the adopted log-normal
distribution is significantly broader than the simulated results (with
the effective gas clumping factor closer to $2.7$ at $z=10$ in the
simulation). Thus our calculation here is conservative, in the sense
that it underestimates true residual absorption hence asymmetry.

\section{Results}
\label{sec:results}

\begin{figure}[t]
\plotone{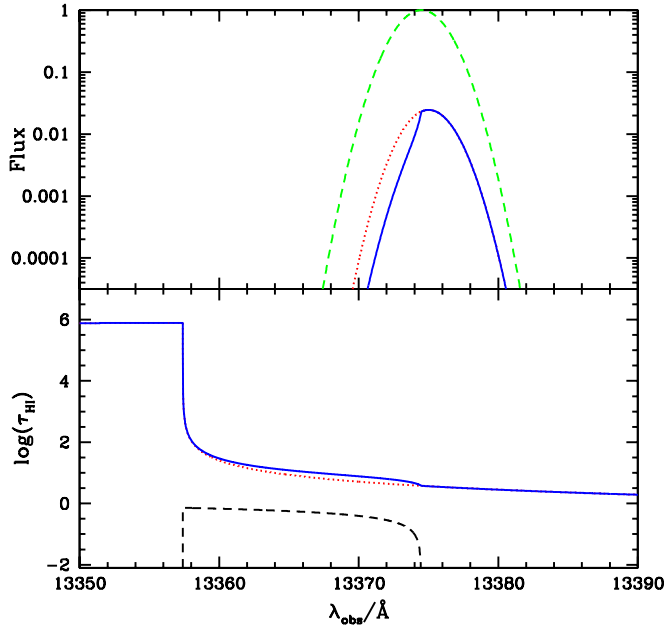}
\caption{The figure shows the effects of absorption by the ambient
IGM on the emitted Ly$\alpha$ line.  The {\it lower panel} shows the
Ly$\alpha$ line opacity from the damping wing of the GP trough of the
IGM (Miralda-Escud\'e 1998) (red dotted curve), and from the neutral
HI within the Str\"omgren sphere (black dashed curve).  Note that for
the latter, we compute the effective opacity $\tau_{\rm eff}=-\ln
\langle \exp(-\tau_{\rm HI})\rangle$, where the brackets denote
averaging over the probability distribution of $\tau_{\rm HI}$. We
assume a log--normal density distribution, with its only free
parameter, $C$ set to 10; and ionization equilibrium within the HII
region, implying $\tau_{\rm HI}\propto \rho^2$.  The blue solid curve
shows the total opacity (sum of GP damping wing and resonant
absorption). The {\it upper panel} shows the intrinsic line emitted by
the galaxy, assumed to be a thermally broadened Gaussian with a width
of $70 {\rm km~s^{-1}}$ (the normalization is arbitrary, shown by the
green dashed curve). The red dotted curve shows the transmitted
profile in the presence of the GP damping wing, but ignoring
absorption from within the HII region.  Note that the line would be
symmetric.  The blue solid curve shows the transmitted profile
including both the GP damping wing and the resonant absorption within
the HII region. The line is asymmetric with $R=0.56$. }
\label{fig:profile} 
\end{figure}

We use a fiducial model to find the attenuation of the Ly$\alpha$ line
as a function of wavelength.  Our model starts with a Gaussian
emission line at $z_s$, and assumes that the source is surrounded by a
spherical Str\"omgren sphere that propagated into the IGM with a
uniform neutral fraction $x_{\rm HI}$.  The fiducial model we adopt
has the following parameters: $z_s$ = 10.00175 (source redshift); SFR
= 3 \msun/yr (or $\dot N_\gamma = 1.1\times 10^{53}$ ionizing photons
per second, corresponding to a usual Salpeter IMF); $\Delta v_i = 70$
km/s (intrinsic linewidth, equivalent to FWHM$=4.4$\AA); $f_{\rm esc}$
= 0.50 (escape fraction of ionizing radiation); $t_s$ = $10^8$ yrs
(age of the source); $\Delta v = 0 $ km/s (velocity offset of line
relative to its surrounding absorbing gas); $x_{\rm HI} = 1$ (neutral
fraction in IGM outside HII region); and a log-normal density PDF with
the chosen clumping factor $C=10$.

Our results for the resulting effective optical depth and transmitted
line profile in the fiducial model are shown in
Figure~\ref{fig:profile}. The key output parameters can be summarized
as follows: $R_s = 0.28$ Mpc physical (3.1 Mpc comoving), $F/F_0 0.02$ (attenuation of total line flux, not including the factor of
($1-f_{\rm esc}$) discussed below), FWHM = 2.3\AA\ of the transmitted
line and $R$ = 0.56.

The attenuation of the line is large, but within the extreme of the
limits quoted by P04 (SFR=0.8 from line, SFR=75 from UV, implying a
factor of 94 attenuation overall).  Note when comparing our results to
the ratio of the SFR inferred from the line and the continuum, the
line attenuation has to be multiplied by the fraction ($1-f_{\rm
esc}$) of the ionizing photons that do not escape from the galaxy and
hence contribute to powering the Ly$\alpha$ line (on the other hand, a
small $f_{\rm esc}$ will result in a small Str\"omgren sphere and
large attenuation; see Haiman 2002 for further discussion). We find
that the product of the attenuation times [$1-f_{\rm esc}$] is
maximized for $f_{\rm esc}=0.5$.  Overall, in the model shown in
Figure 1, the SFR as estimated from the line strength would be a
factor of $0.5\times 0.02=0.01$ of the true value. The linewidth is
reduced by $\sim 50\%$ to $2.3$\AA, but is consistent with the
observed width.  However, {\it the asymmetry is significant}:
$R=0.56$, in clear conflict with the observed value $R=1.12\pm 0.05$
obtained above.

The central point is that while most of the overall line attenuation
is due to the damping wing (the dotted curve in
Figure~\ref{fig:profile}), the asymmetry is caused by the resonant HI
in the HII region (dashed curve in Figure~\ref{fig:profile}), because
the former runs much more smoothly across the central region of the
Ly$\alpha$ line than the latter, in this case with $\Delta v=0$.

There are several ways, in principle, to make the transmitted
Ly$\alpha$ line profile symmetric, to be in accord with the observed
one.  (1) The intergalactic medium is highly ionized by a strong
background flux that reduces the neutral fraction even near the
galaxy. We find that a neutral fraction of $x_{\rm HI}\lsim 1.3\times
10^{-7}$ would be required to produce line with a symmetry parameter
of $R>0.9$.  This would require an unrealistically large ionizing
background of $\Gamma_{12}\sim 1000~{\rm s^{-1}}$.  (2) The intrinsic
width of the Ly$\alpha$ emission line is large.  We find that the
width would have to exceed $\sim 1000$ km/s.  This is in clear
conflict with the observed linewidth of $< 200$ km/s.  (3) The star
formation rate of the galaxy is high, reducing the neutral fraction
inside the HII region. We find that a SFR of at least a factor of
$300$ higher than we used in the fiducial model would be required,
which is unrealistically high compared to the values inferred by P04.
(4) The emitting galaxy happens to sit inside a large HII region
produced by other sources, which are not detected.  None of these
options appears to be physically plausible, except (4), which may
require a more detailed discussion.  For (4) a quasar seems unlikely,
because it would be easily detectable even without gravitational
lensing magnification.  However, a group of strongly clustered small
galaxies around this emitting galaxy which collectively amount to a
luminosity that is more than $300$ times the assumed luminosity of the
detected galaxy could restore the symmetry of the Ly$\alpha$ emission
line.  This would require to pack all of these galaxies into a region
of comoving size of $\le 150$kpc.  While not impossible, this seems
unlikely, because these putative galaxies would be nearly touching one
another, and also because more than one of these galaxies would be
expected to lie close enough to the lensing caustic to be detectable.

\begin{figure}[t]
\plotone{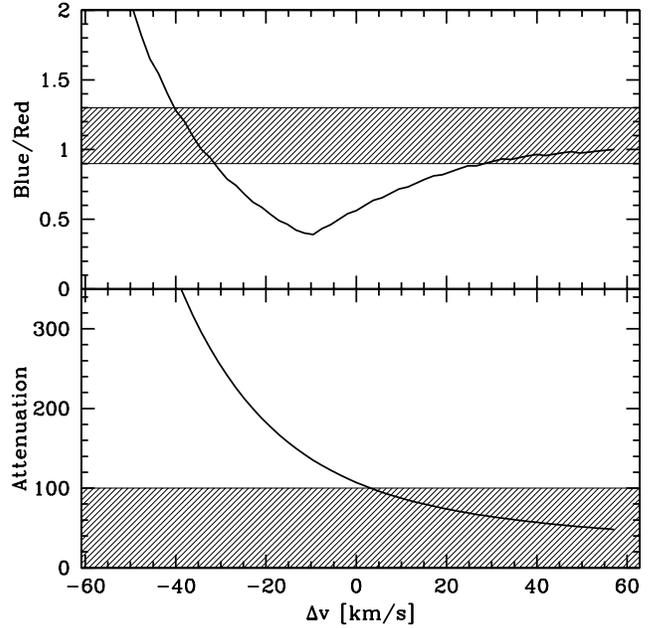}
\caption{This figure shows the apparent line asymmetry (defined as the
ratio of fluxed on the blue and red side of the apparent line), and
the total line attenuation (defined as the ratio of the total
transmitted flux and the emitted flux) as a function of the assumed
velocity offset of the Ly$\alpha$ emitting gas relative to the Hubble
flow.  The cross-shaded regions in both panels indicate permitted
regions, given the observed line profile (top panel) and the observed
Ly$\alpha$ flux to continuum flux ratio (bottom).  Note that the line
can be symmetrized by either a negative or positive velocity.  For
positive velocities (representing extra redshift), the shift needed is
approximately half of the apparent line--width -- to move the line out
of the resonant absorption.  Negative velocities with a similar
magnitude also symmetrize the line.  This is because the line is more
heavily absorbed, and the apparent peak then corresponds to a
wavelength on the red wing of the intrinsic line, with the peak of the
intrinsic line yielding a substantial blue wing for the transmitted
line. However, the overall attenuation of the line in this case is too
large for the parameters of the Pello et al. (2004) source. }
\label{fig:voffset} 
\end{figure}

We here suggest an alternative, physically compelling possibility:
that is, that either the emitting galaxy or the emitting gas in the
galaxy is receding with a velocity relative to the surrounding
absorbing gas in the HII region. Figure~\ref{fig:voffset} shows the
flux ratio and attenuation as a function of the assumed recessional
velocity.  We see that both the observed line profile and line
attenuation can be made to match the observation, if the recessional
velocity is at least $35$km/s.  Essentially, the Ly$\alpha$ emitting
gas needs to recede with a velocity that is comparable to the half
width of the Ly$\alpha$ emission line in order to escape from the
residual absorption inside the Str\"omgren sphere for the blue branch.

Let us now examine the possibility of having such a recessional
velocity.  First, the possibility that the Ly$\alpha$ emitting gas in
galaxy has a recessional velocity of $>35$km/s relative to the
surrounding gas.  Such a velocity may be produced by outflowing gas
powered by supernovae, which is ubiquitously seen at lower redshift
(Shapley \etal\ 2003).  It would be natural to suppose that such a
high--redshift galaxy is also capable of blowing winds at $\ge
35$km/s.  However, this solution is probably not viable, because
outflowing gas is expected to always result in asymmetric intrinsic
Ly$\alpha$ lines with more flux on the red side prior to additional
scattering by IGM, as confirmed by Lyman Break Galaxies (Shapley
\etal\ 2003) at $z=3-4$, which are known to launch winds at speeds
comparable to, or exceeding the linewidths.  Higher redshift
Ly$\alpha$ emitters at $z=5-6$ all seem to show asymmetric Ly$\alpha$
line profiles (Rhoads \etal\ 2003), although the existence of winds is
not yet observed.

Second, we consider the possibility that the galaxy itself has a
recessional velocity of $>35$km/s relative to the ambient IGM.  Using
linear theory, we find that the {\it r.m.s.}  bulk velocity of a
sphere of radius of $0.5$Mpc is $65$km/s at $z=10$ in the standard
LCDM model (Spergel \etal\ 2003), whereas the velocity dispersion
inside such a sphere is only $6.6$km/s, which is much smaller than the
required $35$km/s relative velocity between the galaxy and the
surrounding gas.  Note that most of the absorbing gas causing the
asymmetry arises inside such a sphere (which is about a quarter of the
Str\"omgren sphere).  Therefore, while the galaxy and its surrounding
gas may be moving together at a significant peculiar velocity, a
recessional velocity of $35$km/s of the galaxy relative to its
surrounding absorbing gas is unlikely.  However, another possibility
is that the detected galaxy is a member of a larger, group of
galaxies, which have formed a non--linear system, with a velocity
dispersion of $\ge 35$km/s.  In that case, the galaxy can be moving at
$\ge 35$km/s relative to the surrounding gas.  In addition, the
presence of other, undetected galaxies would somewhat further help
create a larger HII region and hence reduce the asymmetry.  This, in
fact, seems to us a compelling solution, in that it could also explain
why the lower--redshift galaxies do {\it not} have symmetric lines
(they are not part of large enough virialized systems).

\begin{figure}[t]
\plotone{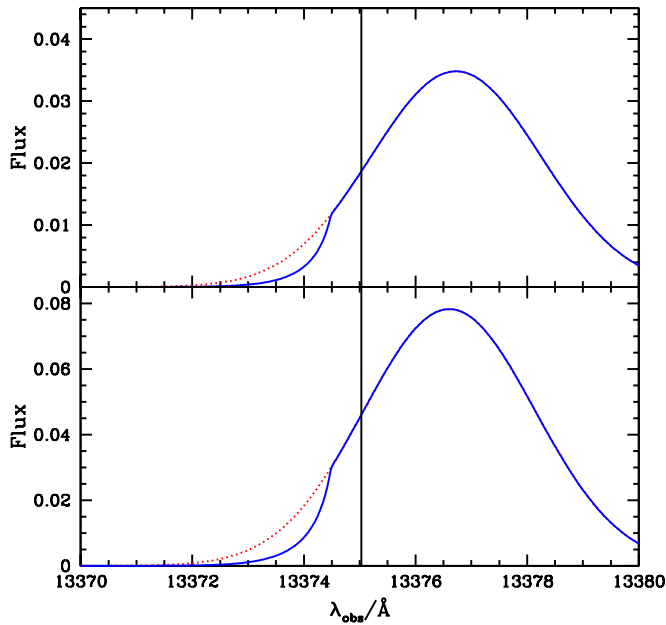}
\caption{The top panel 
shows the case with $x_{\rm HI}=1$ and $\Delta v=35$km/s, and the bottom
panel with $x_{\rm HI}=0.8$ and $\Delta v=35$km/s. In each panel the dotted
curve shows the Ly$\alpha$ emission line with only the damping wing
due to the IGM, whereas the solid curve includes absorption by both
damping wing and residual neutral hydrogen inside the Str\"omgren
sphere.  The total transmission is now reduced by a factor of 60 (top
panel) and 30 (bottom panel) as opposed to 100 in the $\Delta v=0$
model (Figures 1,2).  We further note that for $x_{\rm HI}=(0.6,0.17)$ the
attenuation becomes (13,3), respectively (not shown in the figure).  }
\label{fig:voffset1} 
\end{figure}

The need for this recessional velocity from the line symmetry alone
raises the question: Can we place interesting constraints on the
ionization state of IGM?  Figure~\ref{fig:voffset1} shows the emission
line profile for two cases with different neutral fractions for the
IGM.  While the adopted $f_{esc}=0.5$ is close to maximizing the
transmitted line flux, we note that the Ly$\alpha$ line may have
suffered additional obscuration by, for example, dust in the emitting
galaxy.  The combined uncertainty in $f_{esc}$ and intrinsic
absorption weakens the constraint on $x_{\rm HI}$.  With $x_{\rm
HI}=1$ we find that the overall attenuation due to combined IGM and
residual Ly$\alpha$ scattering is $\le 70$ (for $\Delta v\ge 35$km/s),
which is consistent the upper limit of $100$ of P04.  An $80\%$
neutral IGM produces a total attenuation (including the factor of
$1-f_{\rm esc}=0.5$) of $\le 30$, reasonably close to the mid--range
of $40$ quoted by P04.  However, if the intrinsic absorption including
dust were $80\%$, and if the Ly$\alpha$ line to continuum ratio is at
the low end of range ($0.05$) quoted by P04, then $x_{\rm HI}\le 0.2$
would be preferred.

In conclusion, we find that if one allows a total attenuation of $60$,
then no strong constraint can be placed on $x_{\rm HI}$.  Tighter
constraints can only be made possible with (1) more accurate
calibrations of star formation rates using Ly$\alpha$ emission and UV
continuum, and/or (2) a greatly improved knowledge of the intrinsic
absorption of Ly$\alpha$ emission and UV continuum.  For example, if
one can independently put a upper bound on the combined absorption by
all other sources on the Ly$\alpha$ line, one may be able to place a
lower bound on $x_{\rm HI}$, in addition to an upper bound, given that
the observed line to continuum ratio is quite small.

\section{Discussion}
\label{sec:conclude}

Following the announcement of the $z=10$ candidate galaxy, several
recent works (Loeb, Barkana, \& Hernquist 2004; Ricotti \etal\ 2004;
Gnedin \& Prada 2004) have considered constraints on the neutral
fraction of the IGM.  In particular, allowing for a maximum
attenuation factor of $40$, Loeb \etal (2004) find a mild constraint
($x_{\rm HI}<0.4$) and, allowing for a smaller maximum attenuation
factor of $15$, Ricotti etal (2004) find a stronger constraint that is
alleviated to allow a neutral universe only if $R_s>$5 (comoving) Mpc.
Our results here appear to be consistent with these findings.  Gnedin
\& Prada (2004) have emphasized that a fraction of Ly$\alpha$ galaxies
at $z=10$ could escape strong attenuation due to variations along our
line of sight in the shape and size of HII regions that surround
individual sources at $z=10$.  In our treatment, we have utilized both
the shape and attenuation of the observed line, and we reach a unique
set of conclusions.  In particular, we find that the symmetry of the
line requires that the galaxy has a recessional velocity of $\ge
35$km/s relative to the surrounding gas.  Consequently, we find that
no strong constraint can be placed on $x_H$ at $z=10$ given the
uncertainties.

If our interpretation of the Ly$\alpha$ e mission line profile is
correct, we can deduce that at any redshift there should be a trend:
the fraction of symmetric Ly$\alpha$ emission lines should decrease
with increasing Ly$\alpha$ emission line width.  In addition, bright
Ly$\alpha$ emitters with symmetric profiles may be signposts of groups
and clusters of galaxies, within which they can acquire velocities
comparable to to larger than their linewidths.  We also note that a
velocity dispersion of, say, 65 km/s for the larger group of galaxies
implies a common dark halo mass of $2\times10^{10}~{\rm M_\odot}$ at
$z=10$, and we find, using the halo mass function from Jenkins \etal\
(2001), that there should indeed be $\sim 1$ such halos in the
comoving volume of $\sim 20 $Mpc$^3$ probed by P04 between $9<z<11$.

The expected surface density of dark matter halos between redshifts
$z=9-11$ with mass $5\times 10^8$\msun (the minimum halo mass for the
galaxy itself inferred by P04) is estimated to be $\sim
0.03$arcsec$^{-2}$.  Thus, it should not be a great surprise to find
one galaxy strongly lensed in a targeted search behind a rich cluster
whose Einstein ring size is of order of several arcsec.  However, the
inferred large intrinsic abundance of galaxies is still surprising
when the implied near--IR counts are compared to observations (Ricotti
\etal\ 2004).  It will be highly desirable to enlarge the sample of
such galaxies in the future, for a better estimate of their space
density, as well as to acquire better statistics on constraints for
the neutral fraction in the IGM.

\section{Conclusions}

We considered the implications of the detection of the symmetric,
highly attenuated Ly$\alpha$ emission line from a candidate $z=10$
galaxy.  We find that the observed symmetry of the Ly$\alpha$ emission
line can be accounted for if the emitting galaxy is receding relative
to the surrounding absorbing gas by a velocity of at least $35$km/s.
Such a relative velocity is mostly plausibly achieved if this detected
galaxy is a member of a larger system with a velocity dispersion in
excess of $35$km/s.  Thus, while the difficulties and challenges
associated with such observations are formidable, it is not a great
surprise, in principle, to be able to detect galaxies with such
Ly$\alpha$ emission lines at high redshift.  However, with the
required recessional velocity, this galaxy does not place a strong
constraint on the ionization state of the IGM, given various
uncertainties in the current data and lack of handle of the intrinsic
absorption. A fully neutral universe, while not preferred, is still
consistent with the observation.  

A moderate increase in the sample
size of such high redshift galaxies will be highly valuable in
statistical inferences for the ionization state of the IGM, based on
the systematic dependence of the line properties on redshift and
luminosity (Haiman 2002; Rhoads \& Malhotra 2002).
More urgent in the short term is to obtain a higher quality
spectrum to better characterize the line profile.

\acknowledgements{We thank James Rhoads for useful conversations, and
the authors of P04 for providing the line profile and instrument
response in electronic form. We gratefully acknowledge financial
support by NSF through grants AST-03-07200 and AST-03-07291 (to Z.H.)
and AST-0206299 (to R.C.), and by NASA through grant NAG5-13381 (to
R.C.).}

\vspace{-2\baselineskip}

\end{document}